\documentclass[12pt,preprint]{aastex}
\pdfoutput=1 
\newcommand{\HI}{H~{\sc i}} 

\newcommand{\kms}{${\rm km~s^{-1}}$}

\slugcomment{{\em Accepted for publication in ApJ}}
\shortauthors{McCLURE-GRIFFITHS \& DICKEY} 
\shorttitle{Milky Way Kinematics at the First Quadrant Subcentral Point}
\begin{document} 

\title{Milky Way Kinematics. II.  A uniform inner Galaxy H~{\sc i} terminal velocity
  curve}

\author{N.\ M.\ McClure-Griffiths\altaffilmark{1} 
and John M.\ Dickey\altaffilmark{2}} 

\altaffiltext{1}{Research School of Astronomy and Astrophysics, 
  Australian National University, Canberra ACT 2611, Australia; naomi.mcclure-griffiths@anu.edu.au}

\altaffiltext{2}{School of Mathematics and Physics, University of
  Tasmania, Hobart TAS 7001, Australia;
  john.dickey@utas.edu.au}

%--------------------------------------------
\begin{abstract}
%-------------------------------------------
  Using atomic hydrogen (\HI) data from the VLA Galactic
  Plane Survey we measure the \HI\ terminal velocity as a
  function of longitude for the first quadrant of the Milky Way.  We 
  use these data, together with our previous work on the fourth
  Galactic quadrant, to produce a densely sampled, uniformly measured,
  rotation curve of the Northern and Southern Milky Way between
  $3~{\rm kpc} < R < 8~{\rm kpc}$. We determine a new joint rotation
  curve fit for the first and fourth quadrants, which is consistent with the
  fit we published in McClure-Griffiths \& Dickey (2007) and can be
  used for estimating kinematic distances interior to the solar
  circle.   Structure in the rotation curves is now exquisitely well defined, showing significant velocity structure on lengths of $\sim 200$ pc,  which is much greater than the spatial resolution of the  rotation curve. Furthermore, the shape of the rotation curves for the first and fourth quadrants, even after subtraction of a circular rotation fit shows a surprising degree of correlation  with a roughly sinusoidal pattern between $4.2 < R < 7$ kpc.  
\end{abstract}

%----------------------------------------------------------
\section{Introduction}
\label{sec:intro}
%----------------------------------------------------------
Atomic hydrogen (\HI) emission, as traced by the $\lambda=21$ cm line,
is ubiquitous throughout the Milky Way disk.  As such it is an
excellent tracer of gas kinematics and dynamics in the disk.  There is
a long and fruitful history of using measurements of the \HI\ line to
measure the rotation curve of the Milky Way and to infer the Galaxy's
structure
\citep[e.g.][]{kwee54,kerr62,shane66,burton78,levine06,sofue09}.  The
departures from circular rotation, observed as the various bumps and
wiggles in the rotation curve, have been used to infer the mass structure of
the Galaxy using spiral density-wave theories \citep{lin69,roberts69}
or more recently through comparison with simulations \citep{chemin15}
and mass discrepancy acceleration modelling \citep{mcgaugh16}.
Fundamentally, accurate knowledge of the rotation curve is required
for any attempt to model the surface density profile of a galaxy and
the Milky Way is no exception.  In addition, the Milky Way rotation curve is used extensively for kinematic distance estimates based on the observed radial velocity of an object or associated gas. 

Much of the work on the rotation curve interior to the Solar circle ($R<R_0$) is based on
measurements of the \HI\ or CO terminal-velocity.  Inside the solar
circle each line of sight has a location where it is tangent
to a circular orbit about the Galactic center.  At this location,
known as the tangent point, the projection of Galactic rotational
velocity, $V_{\theta}(R)$ onto the Local Standard of Rest (LSR) is greatest, and the measured LSR velocity is
called the terminal velocity.  At the tangent point all radial or
vertical velocities are projected perpendicular to the LSR and the distance from the Galactic center is geometrically
defined at $b=0\arcdeg$ as $R_t=R_0\sin(l)$.  Assuming that azimuthal
streaming motions are not too large (see however, Chemin et al.\
2015), the maximum velocity of \HI\ emission can be equated to the
terminal velocity, $V_t$, and assigned to a galactocentric radius,
$R_t$.

This is the second paper in a series about the kinematics of the inner
Milky Way.  In the first paper, \citet{mcgriff07}, hereafter Paper I,
we developed a new technique for measuring the \HI\ terminal velocity
curve.  In that paper we discussed the technique extensively and
produced a table of \HI\ terminal velocities sampled every four
arcminutes for the fourth Galactic quadrant using \HI\ data from the
Southern Galactic Plane Survey \citep[SGPS;][]{mcgriff05}.  In this
companion paper we employ the same technique to
measure the \HI\ terminal velocities for the first Galactic quadrant
from the VLA Galactic Plane Survey \citep[VGPS;][]{stil06}.  Together
these terminal velocity curves give a uniform, densely sampled
measurement of the \HI\ rotation curve of the Milky Way between
$3~{\rm kpc} \leq R \leq 8~{\rm kpc}$.  This paper is organized as
follows: in \S \ref{sec:measure} we review the key aspects of the
technique described in Paper I and apply this technique to the VGPS
data.  In \S \ref{sec:termvel} we present the first quadrant terminal
velocity curve, comparing with the CO terminal velocities from
\citet{clemens85} and with the fourth quadrant curve. In \S
\ref{sec:rotcurve} we show both the first and fourth quadrant rotation
curves together with a variety of standard rotation curve fits and
discuss the curves as an ensemble.

%----------------------------------------------------------
\section{Data}
\label{sec:obs}
%----------------------------------------------------------
The \HI\ data used here are from the VLA Galactic Plane Survey
\citep[VGPS;][]{stil06}.  The VGPS is a survey of 21 cm continuum and
\HI\ emission and absorption obtained with the Very Large Array in its
D-array configuration and combined with single dish data from the NRAO
Green Bank Telescope.\footnote{The National Radio Astronomy
  Observatory is a facility of the National Science Foundation
  operated under cooperative agreement by Associated Universities,
  Inc.}  The survey covers Galactic longitudes $l=18\arcdeg$ to
$l=67\arcdeg$ with a varying latitude range from $|b| <1.3\arcdeg$ to
$|b|<2.3\arcdeg$.  The \HI\ spectral line data have an angular
resolution of $1\arcmin$ with rms noise of 2 K of brightness
temperature, $T_b$, in each $0.83~{\rm km~s^{-1}}$ velocity channel.
Here we use the \HI\ data, as well as the continuum data, in the
latitude range $|b|<0.5\arcdeg$.

We compare the first quadrant terminal velocity data with terminal velocity data of the fourth
quadrant from the Southern Galactic Plane Survey, as presented in
Paper I.  The noise and resolution characteristics of the SGPS and
VGPS are comparable, leading to a very uniform dataset across the
first and fourth quadrants.
%----------------------------------------------------------
\section{Measuring the Terminal Velocity}
\label{sec:measure}
%----------------------------------------------------------
For all lines of sight near the Galactic plane the circular rotation
of the Milky Way projects onto the Local Standard of Rest (LSR) as:
\begin{equation}
V_{LSR} = R_{0} \sin{l} \left( \omega - \omega_{0} \right)
\cos{b},
\end{equation}
where $\omega$ is the angular velocity of Galactic rotation at the
point of interest, and $\omega_{0}$ is the corresponding angular
velocity of Galactic rotation at the solar circle ($R = R_{0}$).  In
general we discuss the circular rotation speed, $\Theta(R) = R\omega$.
The IAU recommended values for the distance to the Galactic Center,
$R_0$, and the circular rotation speed at the Sun, $\Theta_0$, are $8.5$
kpc and $ 220~{\rm km~s^{-1}}$, respectively.  Although there is
strong evidence that these should be varied, i.e.\ \citet{reid14}, we
use the IAU values here for consistency with previous work, except
where noted.  By measuring the
terminal velocity, $V_t=V_{LSR}(R)$, at the tangent point, $R_t=R_0\sin{l}$ we can
derive the Galactic rotation curve, $\Theta(R) = |V_t|+\Theta_0|\sin{l}|$, interior to the solar
circle.  In practice, extracting the terminal velocities close to the
solar circle is almost impossible because the circular velocity
projected onto the LSR has nearly zero radial component.  We therefore restrict our
measurements to about $0.92 R_0$ ($l\sim 67\arcdeg$).  

As described in an extensive body of past work on the Milky Way \HI\
rotation curve
\citep[e.g][]{kerr62,shane66,sinha78,celnik79,kulkarni85,malhotra95,levine06},
and summarized in Paper I, if the Milky Way were filled with cold gas
in perfectly circular orbits the \HI\ spectrum would drop abruptly to
zero at the terminal velocity.  In reality, thermal velocity motions,
random gas motions and bulk motions all contribute to extend the \HI\
spectrum beyond the terminal velocity.  The magnitude of thermal gas motions and turbulent
widths are determined from the spectral width beyond the terminal
velocity.

In Paper I we built on the analysis by \citet{celnik79} and \citet{kulkarni85} to cast
the drop-off of the \HI\ profile beyond the terminal velocity in terms
of an error function.  We will not reproduce the full derivation, but
the conceptional point is that for optically thin gas beyond the
terminal velocity, the brightness temperature, $T_b$, as a function of
velocity, $ v$, in each velocity interval $\delta v$ is:
\begin{equation} T_b(v) \ \delta v \ = \  \frac{n_o}{C} \
  \int_{v}^{\infty} \left| \frac{dv^{\prime}}{dr} \right|^{-1}
  f(v-v^{\prime}) dv^{\prime}, 
\label{eq:tb}
\end{equation}
where $n_0$ is the local gas number density in units of
${\rm cm^{-3}}$; $C$ is the familiar constant converting velocity integrated
brightness temperature to column density,
$C=1.823 \times 10^{18}~{\rm cm^{-2}~(K~km~s^{-1})^{-1}}$;
$f(v - v^{\prime})$ is the velocity distribution of HI atoms; and
$\left| dv^{\prime}/dr\right|$ is the magnitude of the velocity
gradient at the position, $r$, along the line of sight,  where the radial velocity is $v^{\prime}$.
\citet{celnik79} showed that for a smooth rotation curve
$\left| dv/dr \right|$ is small and directly proportional to $r$,
taking values of $<3~{\rm km~s^{-1}~kpc^{-1}}$ for Galactic longitudes
$| l| >20\arcdeg$.  We can therefore move the gradient term outside of the
integral and if the velocity distribution function of \HI\ atoms,
$f(v-v^{\prime})$, is Gaussian, then the drop-off in the \HI\ profile
beyond the terminal velocity as shown in equation (\ref{eq:tb}) is an
error function.  In fact, we showed in Paper I that the data were best
fit by a sum of two or three error functions of widths, $\Delta v_i$:
\begin{equation}
  T_b(v) = \sum_{i=1}^N \, a_i\, {\rm erfc}(-1*(v-v_o)/\Delta v_i),
\label{eq:errfunc}
\end{equation}  
where $N$ is the number of error functions. Here the term $v_o$ is the
velocity offset between the terminal velocity and a brightness temperature
threshold, which we use to define a starting place for fitting the
\HI\ spectrum at the terminal velocity.

To find the terminal velocity some authors have defined a brightness
temperature threshold \citep[e.g.][]{malhotra95} or a relative
brightness temperature threshold such as half of the intensity of
the most extreme peak \citep[e.g.][]{sinha78}, whereas \citet{shane66}
and \citet{celnik79} fit a functional form, similar to the error
function described above.  Any threshold method is
subject to errors because both the local density, $n_0$, and the path
length, $\Delta r = \delta v\left| dr/dv\right|$, contribute to the \HI\ brightness
temperature (as shown in eq. \ref{eq:tb}).  In Paper I we found best results
with a functional fitting method that was initiated from a
threshold.  We showed that the threshold was useful to determine a
starting point for the functional fitting, but that by fitting the
spectrum with a sum of two error functions we were able to largely remove
ambiguities caused by the choice of threshold (see \S 3.3 of Paper
I).  

An advantage of using a survey like the VGPS or SGPS for measuring
terminal velocities is that the effects of \HI\ absorption of
continuum emission can be removed from the data.  Absorption can
alter inferred terminal velocity by reducing the
brightness temperature of velocities near the terminal velocity.  The
net effect is a reduction in the terminal velocity  at positions where there is
significant absorption. A clear example was shown in Figure 1 of Paper I.  In our technique we use the
VGPS continuum images to define a mask of pixels to exclude from our
analysis.  Masking at a level of 40 K in the continuum images we
reduce the total number of pixels by $\sim 1$\% but also decrease
the scatter in the terminal velocity measurements by $\sim 3-5$\%.  

Our steps to finding the terminal velocity from the VGPS were
therefore:
\begin{enumerate}
\item Mask \HI\ spectra towards pixels containing continuum emission in excess of 40 K, as
  measured in VGPS continuum images.
\item Find the velocity of the 20 K $T_b$ threshold on latitude
  averages over the range $|b|<0.5\arcdeg$.  We chose a threshold value of 20 K, as
  opposed to 10 K as used in Paper I, because the noise in the VGPS
  \HI\ data is slightly higher, such that the threshold is at $\sim 10\sigma$.
\item Shift all profiles to place the threshold crossing velocity, $v_t$, at a common velocity of $0$ \kms\  to simplify binning with longitude and the functional fitting.
\item Average the shifted spectra in longitude by 3\arcmin\ to improve the signal-to-noise of a given longitude.   
\item Fit  each averaged spectrum with a
  sum of two error functions as defined by Equation \ref{eq:errfunc}
  to determine the true terminal velocity as $V_t = v_t-v_o$.
\end{enumerate}

%----------------------------------------------------------
\section{The Terminal Velocity Curve}
\label{sec:termvel}
\begin{figure}
\centering
\includegraphics[width={\textwidth}]{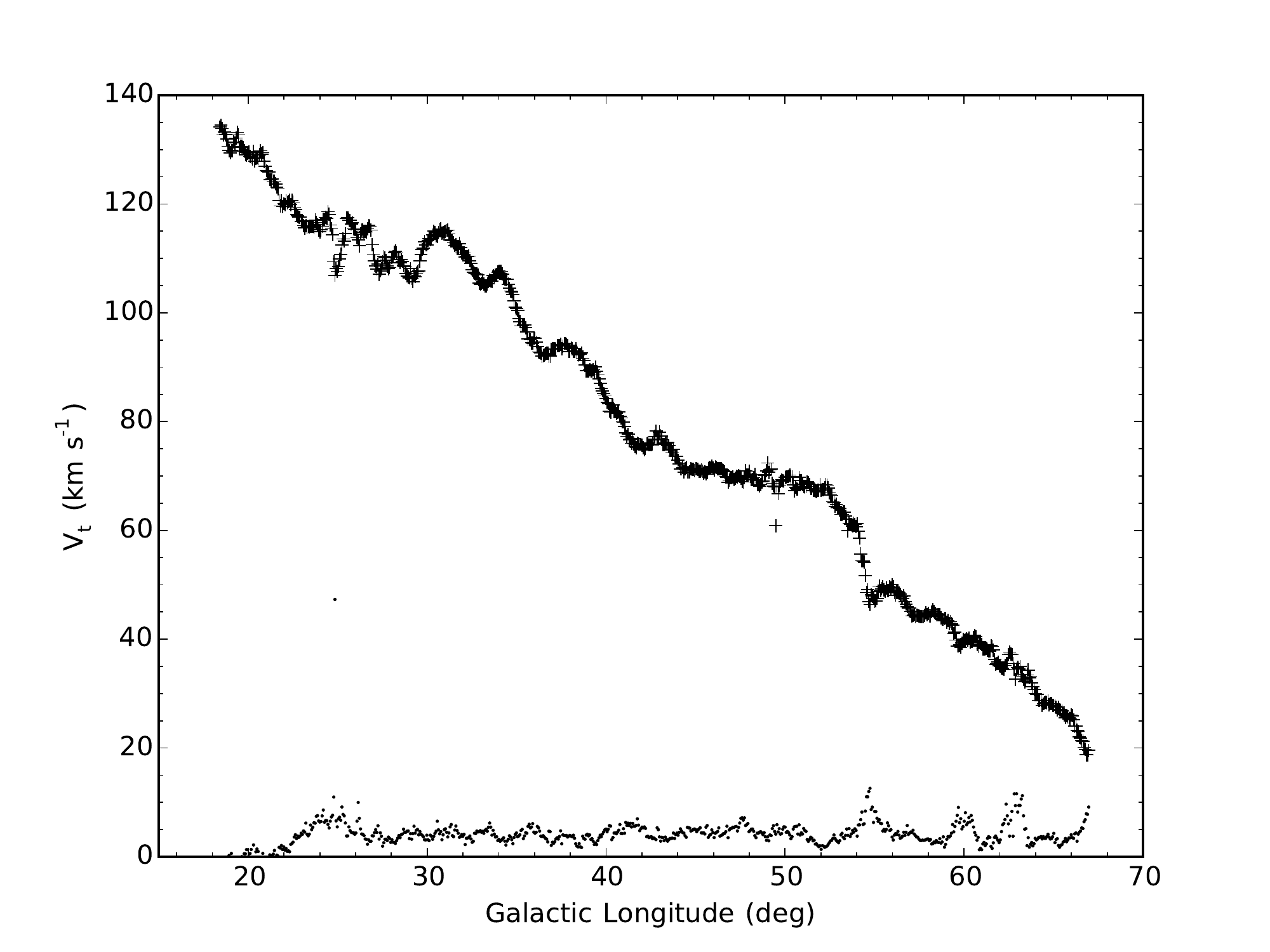}
\caption[]{\HI\ terminal velocities, $V_t$ (crosses), and the velocity offset, $v_o$ (dots), as a
  function of Galactic longitude for the first quadrant extracted from the VGPS.  These data are averaged over 3\arcmin\ in longitude.
\label{fig:vt_vs_l}}
\end{figure}

%----------------------------------------------------------
The first quadrant terminal velocity curve for 960 longitude bins of
width 3\arcmin\ between $l=18\arcdeg$ and $l=67\arcdeg$ is shown
in Figure~\ref{fig:vt_vs_l}.  We also show the offset velocity,
$v_o$, which indicates regions where the terminal velocities are
likely effected by significant local non-circular velocities.  The 
spike in velocity offset at $\l\approx 25\arcdeg$ corresponds to small, anomalous
velocity clouds found by \citet{stil06a}.  In total we found spurious
fits for much less than $1$\% of the fitted spectra.  For the 
analysis that follows we run a median filter of spectral width 3 samples across the $v_o$
values before correcting the data from $v_t$ to $V_t$.  The final terminal velocity values
are provided in tabulated form in Table~\ref{tab:termvels}.

\subsection{Comparison with CO}
The pre-existing most densely sampled terminal velocity curve of the first quadrant
comes from $^{12}$CO J=1-0 data published by \citet{clemens85}.  This curve,
and its associated polynomial fit, has been used for many analyses and kinematic distances
over the past decades.  As a tracer of the terminal velocity CO
has different strengths and weaknesses from \HI.  While \HI\ is
a more ubiquitous tracer, it has large ($\sim 5 - 10$ \kms) velocity linewidths, whereas CO
typically has linewidths of less than $\sim 2$ \kms\ \citep{burton76}.
It has therefore been argued that CO may be a better tracer of the
terminal velocity than \HI.  However, in Paper I we compared \HI\ and
CO for the fourth quadrant and showed that they were remarkably
consistent but with larger scatter in the CO measurements, which we
posited was because the low filling factor of CO led the CO terminal velocity curve to reflect the individual velocity dispersion of clouds, rather than an
ensemble average as traced by \HI.  The analysis there indicated that
neither tracer was significantly better than the other.

\begin{figure}
\centering
\includegraphics[width={\textwidth}]{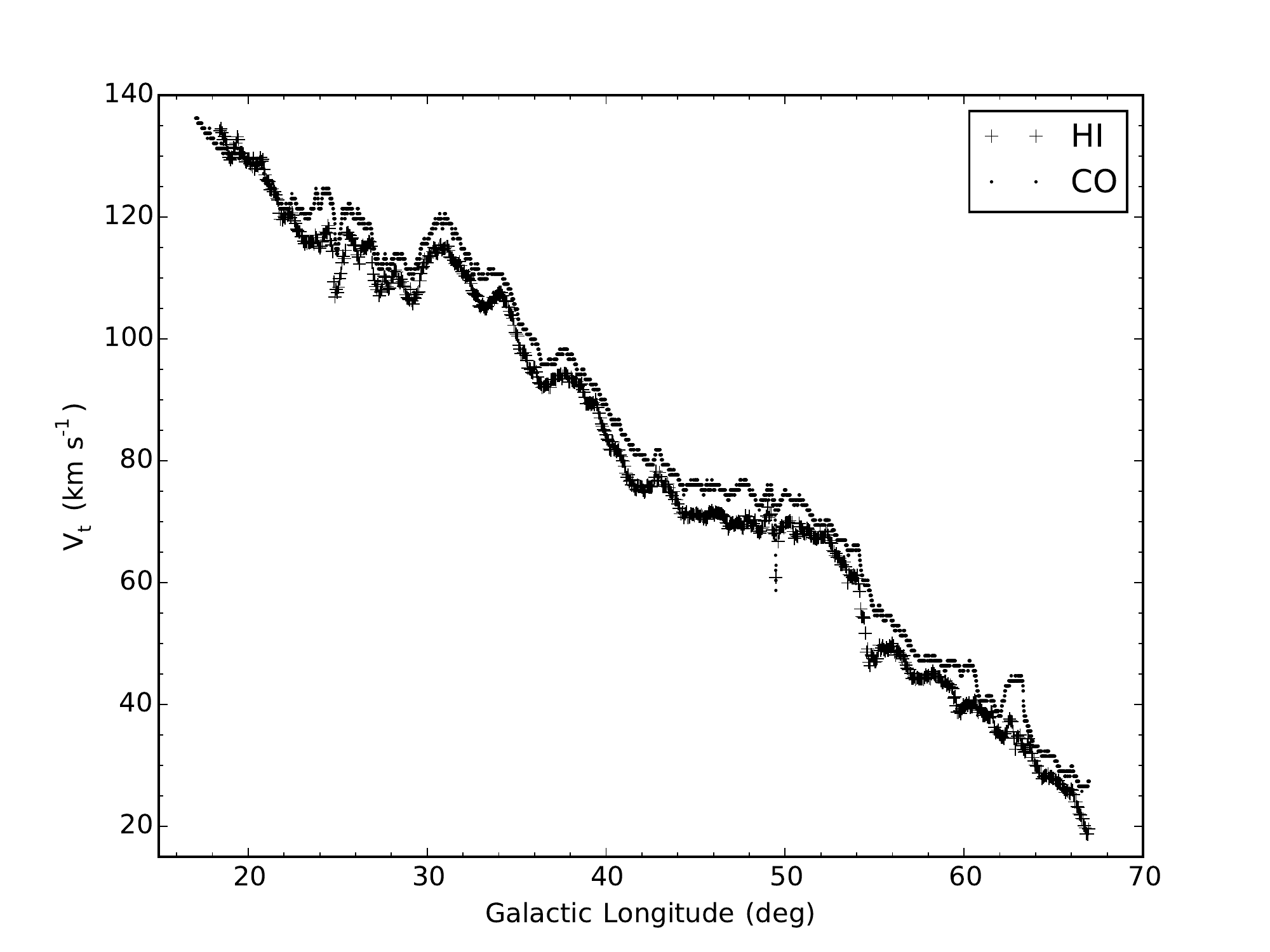}
\caption[]{Comparison of \HI \ (crosses) and CO  (dots) first quadrant terminal
  velocities.  The overall agreement between the two curves is extremely good, but the CO shows a systematic offset towards higher velocities.
\label{fig:vt_hi+co}}
\end{figure}

Figure~\ref{fig:vt_hi+co} shows the \citet{clemens85} curve plotted
with the \HI\ curve determined for the first quadrant.  At all longitudes
$l>22\arcdeg$ there is a systematic offset between the two curves,
with the \citet{clemens85} CO data approximately 7 \kms\ higher than
the \HI.  In deriving the CO terminal velocity curve, \citet{clemens85} effectively used a threshold detection
method (three consecutive channels with emission greater than
$3\sigma$) for the most extreme velocity line component, corrected by
a uniform offset towards permitted velocities (smaller $V_{LSR}$) equivalent to the
half-width of an average CO cloud or $3\pm1.5$ \kms.  In general,
threshold-crossing methods will result in higher terminal velocities than our spectral
fitting method, which locates the inflection point of the spectral
drop-off.  The difference between \citet{clemens85} and our \HI\
results suggests that the average CO cloud 3 \kms\ correction is
insufficient to recover the velocity at the tangent point.  In spite
of the differences in the magnitude of the velocity at a given 
longitude, the shape of the two curves are reassuringly similar.  
Most of the features on scales as small as a few degrees and up to
scales of tens of degrees are apparent in both curves.  This lends
confidence to the veracity of the small-scale features as true
dynamical effects rather than measurement artefacts.  

For a galaxy in purely circular rotation the run of terminal
velocities will be completely linear with $\sin(l)$:
$V_t = v_{circ}(1-\left|\sin l \right|) + v^{\prime}$.  Of course, non-circular
motions induced by spiral arms, the Galactic bar and other discrete
dynamic effects  produce deviations from linearity, but a linear fit with $\sin(l)$ provides a
useful mechanism for comparing data.  Independently fitting the CO and \HI\ terminal velocity curves and then comparing the residuals we find that the scatter in the two curves is nearly identical at $\sim 6$ \kms. 

\subsection{First and Fourth Quadrant Comparison}
\label{subsec:vt_comp}
\begin{figure}
\centering
\includegraphics[width={\textwidth}]{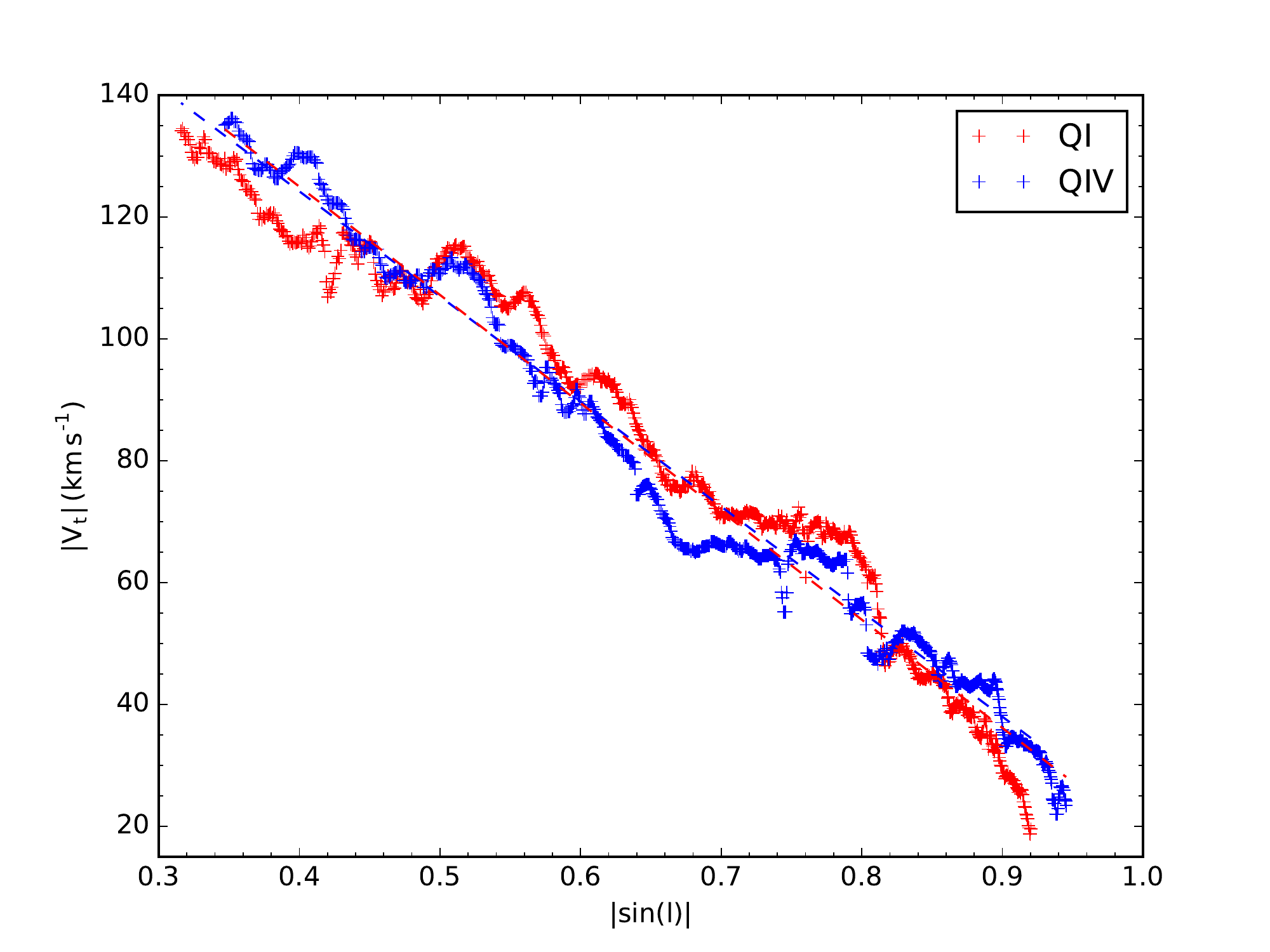}
\caption[]{Comparison of \HI\ terminal velocities for the first
  quadrant (red) and the fourth quadrant (blue). The dotted lines are
  fits to the function $V_t = v_{circ}\, (1- \left|\sin(l)\right|) +
  v^{\prime}$ over the entire longitude range, where the coefficients are
  $v_{circ}=172.5$ \kms\ and $v^{\prime}=20.7$ \kms\   for QI and $v_{circ}=177.7$ \kms\ and $v^{\prime}=18.4$ \kms\  for QIV.  
\label{fig:vt_both}}
\end{figure}

We compare \HI\ terminal velocity curves for the first and fourth quadrants, each plotted versus $|\sin(l)|$, in
Figure~\ref{fig:vt_both}.  The absolute values of the velocities are
plotted for direct comparison.  The two curves are strikingly similar,
despite probing different regions of Galactic structure.  Most
notably, both curves have an abrupt flattening of the slope at
$|\sin(l)|=0.67$ ($|l|\sim 42\arcdeg$) with almost no gradient in velocity 
until $|\sin(l)|=0.67$ ($|l| \sim 52\arcdeg$).  In addition, both curves show an
inflection at $|\sin(l)| = 0.5$ ($|l|\sim 30\arcdeg$).  These features will be
discussed further in the context of the rotation curve below.

The dashed lines in Figure~\ref{fig:vt_both} show independent linear fits  to the run of
terminal velocities with $|\sin(l)|$ in QI and QIV of the form described above ($V_t = v_{circ}(1-\left|\sin(l) \right|) + v^{\prime}$), where the fitted coefficients are $v_{circ}=172.5$ \kms, $v^{\prime}=20.7$ \kms\ and
$v_{circ}=177.7$ \kms, $v^{\prime}=18.4$ \kms\ for QI and QIV, respectively.  It is important to note that because the rotation curve is not necessarily flat over this range of $\sin(l)$, the coefficient $v_{circ}$ does not equate to $\Theta_0$.  However, just as with the comparison of the \HI\ and CO curves, the fit is a useful indicator of the consistency of the two quadrants.  We find that the fits to QI and QIV are consistent to within 2\% indicating the uniformity of the
complete terminal velocity curve presented here.  

\begin{figure}
\centering
\includegraphics[width={\textwidth}]{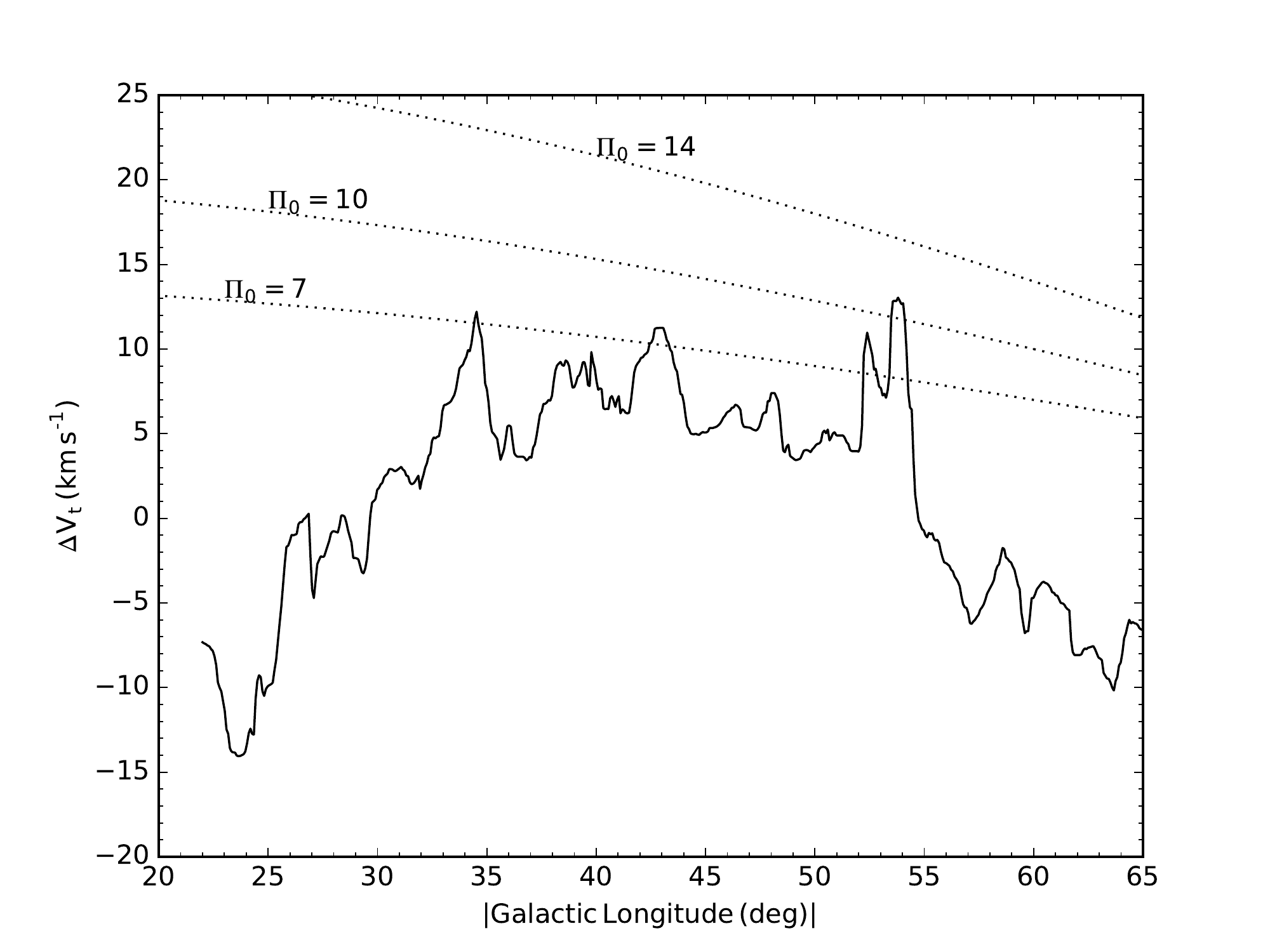}
\caption[]{Difference between QI and QIV terminal velocities as a
  function of Galactic Longitude.    This curve has been smoothed by a Gaussian kernel of width $1\arcdeg$.  The dotted lines show $\Delta v = 2 \Pi_0 \cos l$, for $\Pi_0=7$ \kms, $10$ \kms, and $15$ \kms, as labelled.
  \label{fig:vdiff_l}}
\end{figure}

There are, however, some large-scale offsets between the two curves.  The asymmetry between the the terminal velocity curves in the first and fourth  quadrants has been discussed extensively \citep[e.g.][]{kerr62,grabelsky87,blitz91}.    In Figure~\ref{fig:vdiff_l} we plot a smoothed version of the difference between our QI and QIV terminal velocities as a function of Galactic longitude.  The difference ranges smoothly from $-15$ \kms\ at $|l|\sim 25\arcdeg$ to $10$ \kms\ at $|l|\sim 40\arcdeg$ and back down to $-10$ \kms\ at $|l|\sim 65\arcdeg$. As discussed in \citet{blitz91}, the difference between the terminal velocity curves depends only on the difference in azimuthal velocity of the gas at the tangent points in the two quadrants and any potential outward motion of the LSR. An outward motion of the Sun or the LSR, as posited by \citet{kerr62} and \citet{blitz91},  will give rise to a velocity difference between QI and QIV at all longitudes of the form $\Delta v = 2 \Pi_0 \cos l$, where $\Pi_0$ is the radial velocity  of the LSR. \citet{kerr62} found $\Pi_0 = 7$ \kms, which was consistent with stellar data at the time.  
\citet{blitz91} revisited the asymmetry, recommending a value of $\Pi_0 = 14$ \kms for the magnitude of the radial motion of the LSR.  In Figure~\ref{fig:vdiff_l} we over-plot the functional form, $\Delta v = 2 \Pi_0 \cos l$ for radial velocities of $\Pi_0=7$ \kms, $10$ \kms\ and $14$ \kms.  It is clear that over the range of longitudes measured here   an outward motion of the LSR alone cannot account for velocity difference. To fit the terminal velocity difference curve \citet{blitz91} included a non-axisymmetric velocity field induced by a triaxial spheroidal potential. Alternative fits to the Galactic potential, including those associated with the long bar \citep[e.g.][]{chemin15} or spiral arms \citep[e.g.][]{pettitt14} would also produce a noticeable difference curve.  Fitting the velocity curves to multiple Galactic potentials is clearly an important next step, but beyond the scope of this paper.  

An alternative suggestion to explain only the asymmetry between
the two curves near the Solar circle  is that of streaming motions associated with the Carina spiral arm in QIV at $l\approx295\arcdeg$
\citep{henderson82,grabelsky87}.  In McG07 we compared SGPS QIV values with QI data from \citet{fich89}, which showed the quadrants deviating by about 10 \kms\ from $R = 7 - 8 $ kpc ($|l| \sim 55\arcdeg - 67\arcdeg$). Using the consistently measured data here, and shown in Figure~\ref{fig:vdiff_l}, we see that over the longitude range $|l| \sim 55\arcdeg - 67\arcdeg$ the terminal velocities in QI are $3-9$ \kms\ (mean $5.2$ \kms) lower than QIV, with no indication that the difference is converging towards zero at the Solar circle.

%---------------------------------------------------------------------
\section{Inner Galaxy Rotation Curve}
\label{sec:rotcurve}
%---------------------------------------------------------------------
From the terminal velocities measured here we can calculate the rotation curve for $3 \leq R \leq 8$ kpc
from the circular velocity relation
$\Theta(R)/\Theta_0 = \left |V_t \right|/\Theta_0 + \left| \sin l \right|$ and
assuming $R_t = R_0 \left| \sin l \right|$, where we use IAU standard
values of $R_0 =
8.5$ kpc and $\Theta_0 = 220~{\rm km~s^{-1}}$.
The rotation curves for QI, as calculated in this paper, and QIV, as
calculated in Paper I, are shown in Figure~\ref{fig:rot_fit}.  

\subsection{Fits to the Rotation Curve for \boldmath{$3 \leq R \leq 8$} kpc}
\label{subsubsec:fit}
\begin{figure}
\centering
\includegraphics[width={\textwidth}]{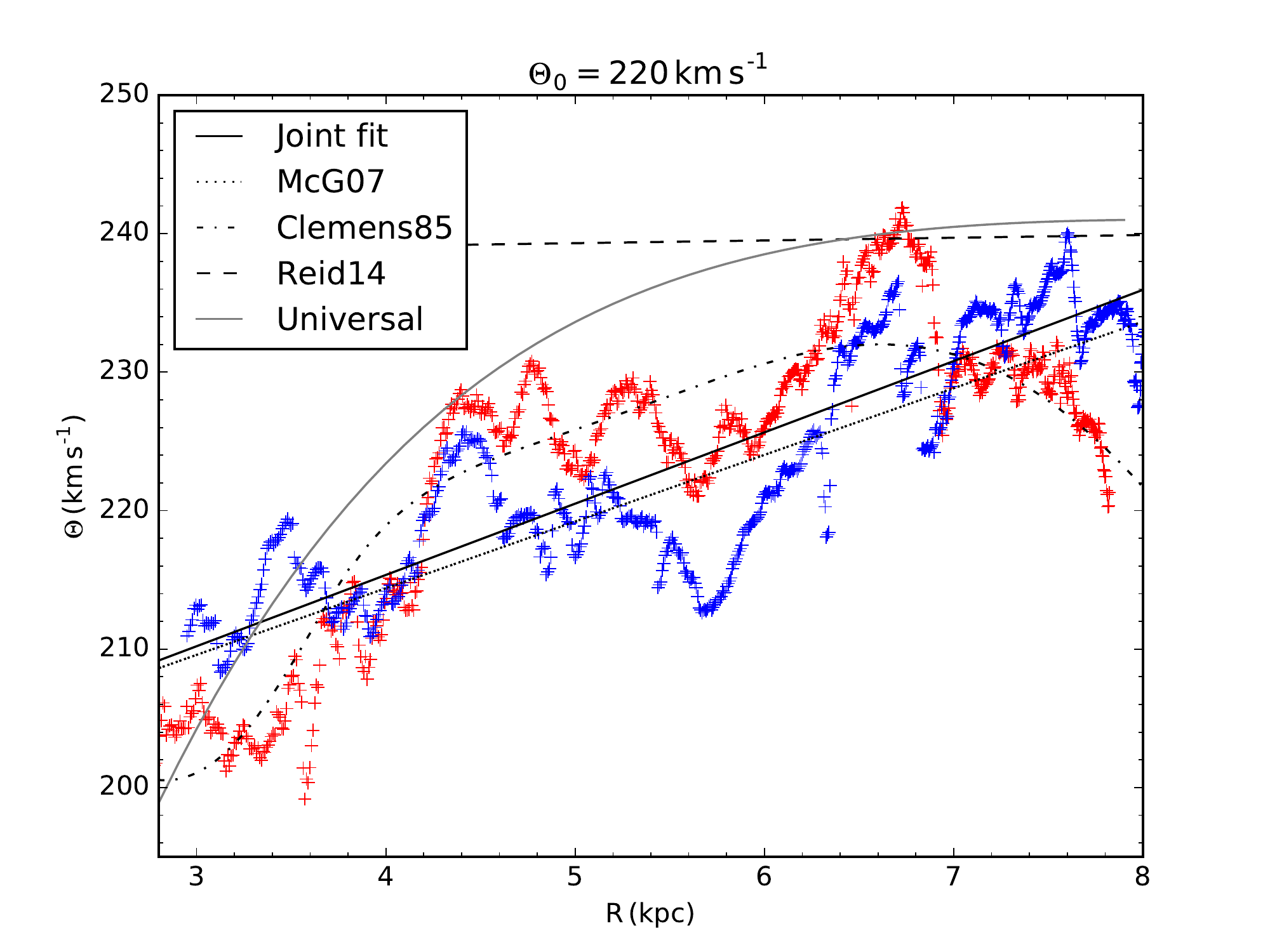}
\caption[]{Rotation curve for the first
  quadrant (red) and the fourth quadrant (blue),  using $\Theta_0=220$ \kms.  Overlaid are the polynomial fit from Paper I (McG07), a new
  joint polynomial fit, $\Theta(R)/\Theta_0 = 0.171 (R/R_0) +
0.889$ and two curves as published in \citet{reid14}.  The first
  of the 
  \citet{reid14} curves is valid for $R>4$ kpc and given by $\Theta(R) = \Theta_0 - d\Theta/dR \,
  (R-R_0)$, where $d\Theta/dR=-0.2$ and $\Theta_0=240$ \kms; and the Reid fit to the  so-called ``Universal''
  rotation curve for spiral galaxies from \citet{persic96}. 
\label{fig:rot_fit}}
\end{figure}

In Paper I we performed a joint linear fit to the QIV data along
with \HI\ rotation curve data of the first quadrant from \citet{fich89}, giving a fit of:
$\Theta(R)/\Theta_0 = 0.186 (R/R_0) + 0.887$.  We repeat that exercise
here substituting the new uniformly measured, high resolution \HI\ data for first
quadrant.  The joint fit is very similar, as can be seen in
Figure~\ref{fig:rot_fit}, $\Theta(R)/\Theta_0 = 0.171 (R/R_0) +
0.889$.  The slightly smaller slope in the new fit is not within
the formal fitting errors of either fit, but is less than 10\% different and
therefore on the order of the error in any given velocity
measurement.   Also shown in this Figure is the often used \citet{clemens85} rotation curve, derived from CO terminal velocities for QI.  While this curve fits well the velocity structure of QI, it incorporates streaming motions that are not general to the inner galaxy rotation curve.

\begin{figure}
\centering
\includegraphics[width={\textwidth}]{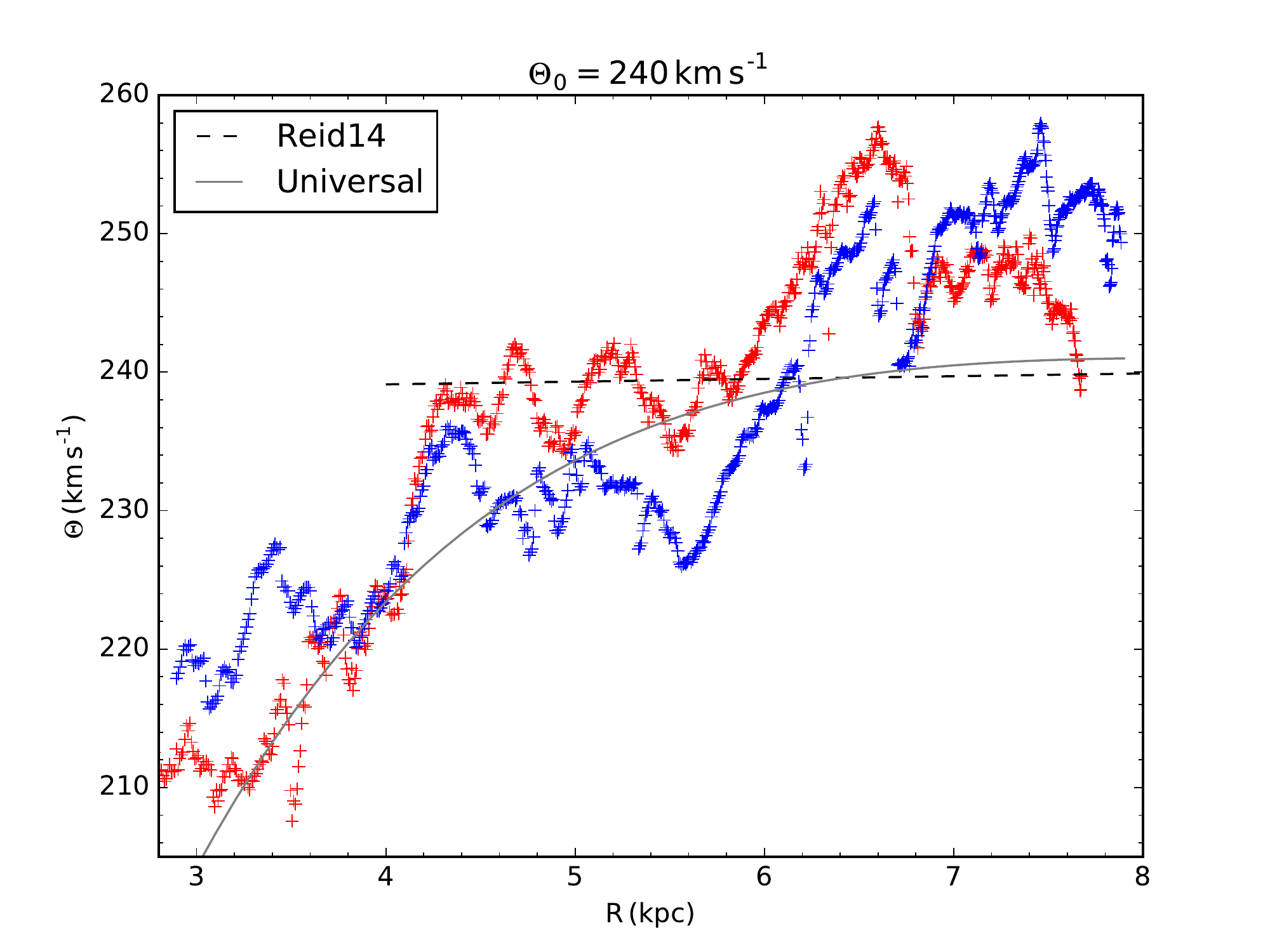}
\caption[]{Our QI (red) and QIV (blue) \HI\ rotation curves calculated assuming $\Theta_0=240$ \kms\ overlaid with the \citet{reid14} rotation curves  shown in Figure~\ref{fig:rot_fit}.
\label{fig:rot_fit_reid}}
\end{figure}

\citet{reid09,reid14} have recently used parallax
measurements of masers to carefully measure the Milky Way rotation
curve and derive Galactic constants, $\Theta_0$ and $R_0$.  In \citet{reid14} they explored several rotation curve fits to their data.  Their simplest fit, which is an excellent match to their data at
$R>4 $ kpc, is given by $\Theta(R) = \Theta_0 - d\Theta/dR\,
  (R-R_0)$, where $d\Theta/dR=-0.2$ and $\Theta_0=241$ \kms.  This produces an almost flat
  rotation curve for large radii, as shown in Figures \ref{fig:rot_fit} \& \ref{fig:rot_fit_reid}.  Including data interior to $R=4$ kpc,
  \citet{reid14} found  that the  so-called ``Universal''
  rotation curve for spiral galaxies from \citet{persic96} captured
  the decrease in velocities at $R<4 $ kpc, while still maintaining the
  nearly flat, though slightly declining, rotation curve for larger
  radii.  Their fit  to the ``Universal'' curve is also shown in Figure~\ref{fig:rot_fit} using their value of $\Theta_0 = 240$ \kms.   Both the \citet{reid14} rotation curves are above our \HI\  velocities calculated from IAU value of $\Theta_0=220$ \kms.  For comparison, in Figure~\ref{fig:rot_fit_reid} we show our \HI\ rotation curves calculated assuming $\Theta_0=240$ \kms\ and $R_0=8.34$ kpc, as in \citet{reid14}.  For these values, the `Universal'' rotation curve gives a reasonable match to the data. Because of the small range of Galactic radii available in our data we  do not perform a new fit with the `Universal'' rotation curve.  Future work should include a
 careful compilation of the extremely accurate maser parallax measurements with these new \HI\ curves
 for a complete `Universal'' Milky Way rotation curve. For
 now we recommend  the joint linear fit as a good estimate of the
 \HI\ rotation curve at $3~{\rm kpc} < R<8$ kpc.

\subsection{Kinematic Distances}

An advantage of our simple linear fit to the rotation curve over the range $R=3$ to $8$ kpc range, i.e.\ $\Theta(R)/\Theta_0  =   k_0 (R/R_0) + k_1 $, where  $k_0 = 0.171$ $k_1 = 0.889$, is that kinematic distances are given by the quadratic equation, based on equation 1 and the law of cosines.  Scaling the observed velocity
of an object of interest, $V_{LSR}$, by $(\Theta_0 \sin{l})$ gives  
dimensionless parameter $x \ \equiv \ V_{LSR}/(\Theta_0   \sin{l})$.  We can therefore define the kinematic distance, $D_k$, in terms of the parameter $x$ as: 
\begin{equation}
\frac{D_k}{R_0}  = \cos{l} \pm  \sqrt{ \frac{0.790}{\left(x \ + \ 0.971\right)^2}  - \sin^2{l}   },
\end{equation}
where the constants are $1 - k_0^2 = 0.971$ and $k_1^2 = 0.790$.  The square root term gives the line of sight distance measured on either side of the subcentral point, whose distance is defined as $d_t = R_0 \cos{l}$.

\subsection{Structure in the Rotation Curve}
\begin{figure}
\centering
\includegraphics[width={\textwidth}]{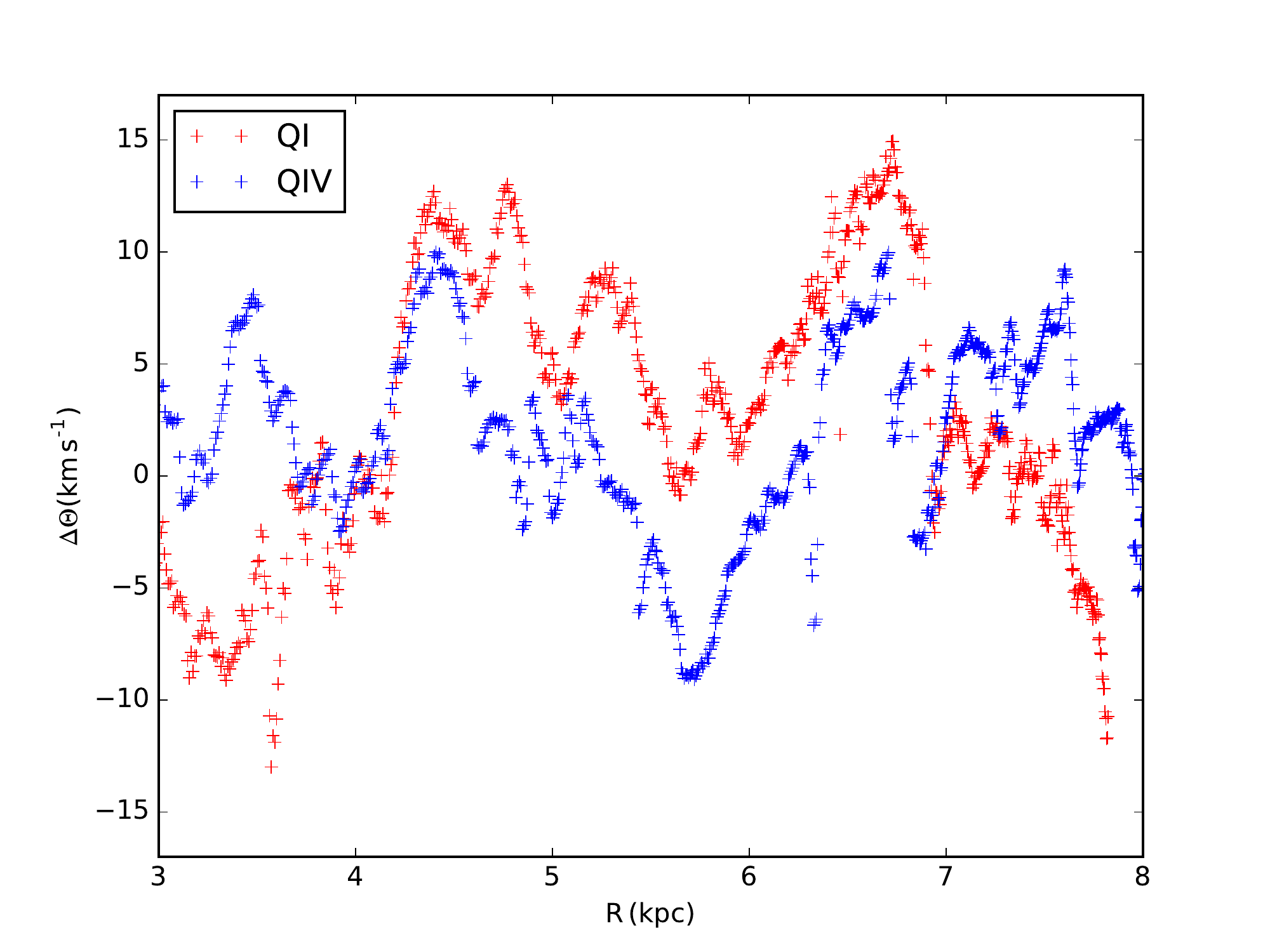}
\caption[]{Rotation curve residuals after subtraction of a joint
  rotation curve fit.  Residuals are shown for QI (red) and QIV (blue)
  after subtraction of  the joint, linear fit  shown in Fig.~\ref{fig:rot_fit}.
\label{fig:resids}}
\end{figure}

Using the joint polynomial fit described above we can subtract the
overall rotation to examine the structure of the residuals in the
rotation curve.  These residuals should reflect bulk non-circular motions in the Galaxy.  The residual curves for both quadrants are shown in
Figure~\ref{fig:resids}.  Once again, QI is shown in red and QIV in
blue.  The values for the two quadrants are extremely close at $R\sim 4$ kpc and for
$R>7$ kpc.  The agreement at $R>7$ kpc is expected as we trace gas towards the cardinal points $l=90\arcdeg$ and $l=270\arcdeg$, where circular rotation projects towards $V_{LSR} = 0$ \kms.  The agreement at $R<4$ kpc is more surprising.  While there are clearly small scale variations with $R$, i.e.\ $\Delta R<500$
pc, the magnitude of those features are generally $\pm 5$ \kms.  The
largest deviations occur in the range $4.2\leq R\leq 7$ kpc.  In this region the two
curves are offset from each other  by $\sim 3 - 9$ \kms.  This region also encompasses some of the salient features
noted in the terminal velocity curve discussed in
\S\ref{subsec:vt_comp}.  Notably the inflection seen in both
quadrants' terminal velocity curves at $|l| \sim 30\arcdeg$ is
observed as the rise in both quadrants' rotation curves at
$R\sim 4.25$ kpc and the flattening of the slope of the terminal
velocity curves between $|l|\sim 42\arcdeg$ and $|l| \sim 52\arcdeg$ are observed
as the smooth rise in the range $R\sim 5.7 -6.7$ kpc.

\subsection{Residual Velocity Field}

The most striking aspect of the residuals shown on Figure~\ref{fig:resids} is the similarity in the saw-tooth patterns seen in both quadrants between $4 < R < 7$ kpc.
In addition, by eye it appears  that there is a great deal of structure in both quadrants that has a
characteristic width of a few hundred pc, even though the resolution of
the survey is much finer than this, typically $\sim$10 pc x $\cos{ l}$.  
Although the effects of \HI\ absorption toward compact continuum sources have
been removed, there are a few very narrow features on Figure~\ref{fig:resids} that correspond
to local velocity perturbations, most of the structure in both residual curves is much
broader in $R$.

\begin{figure}
\centering
\includegraphics[width={\textwidth}]{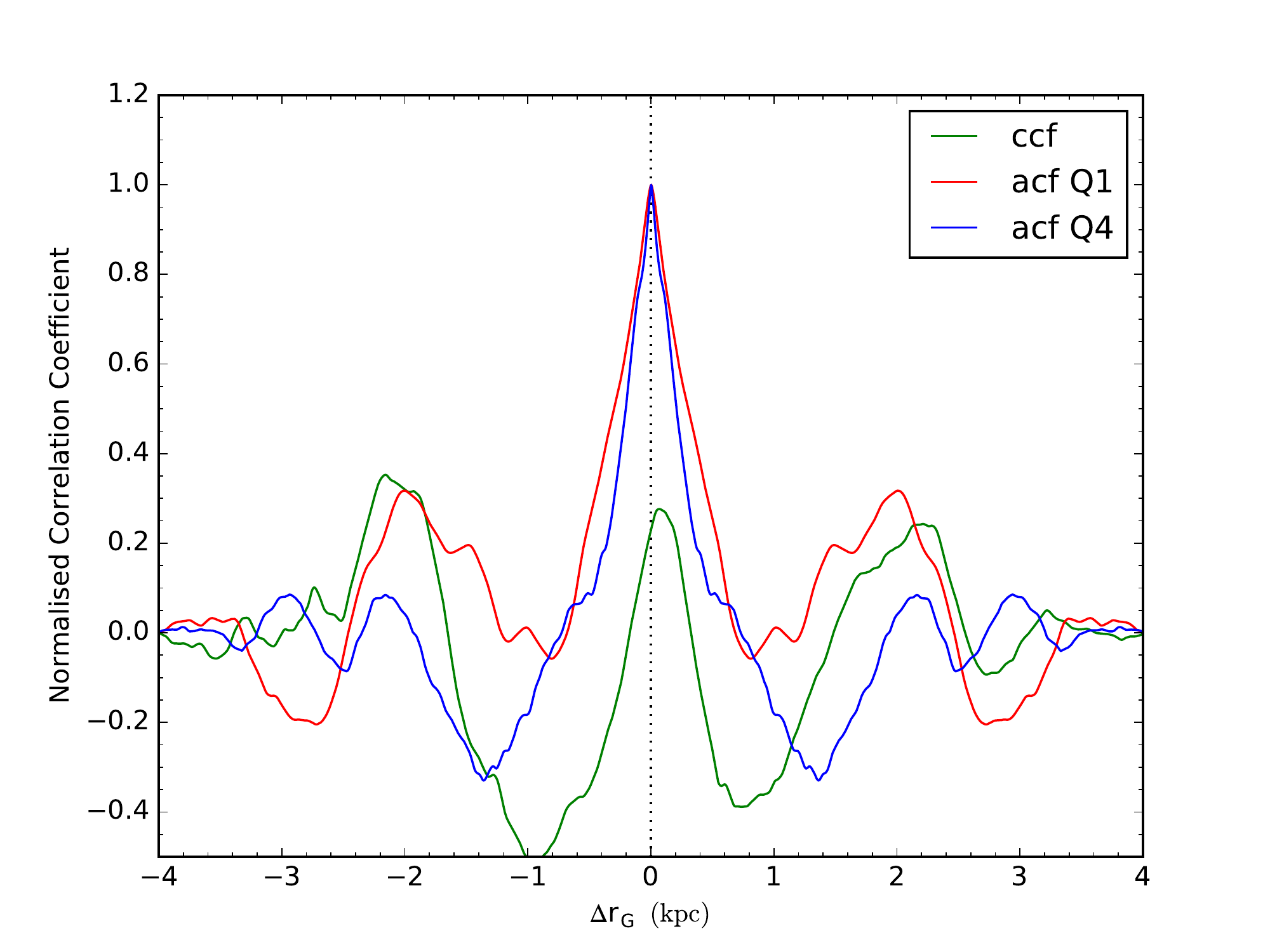}
\caption[]{Correlation functions of the rotational velocity residuals after subtraction of  a linear fit, as shown in Figure ~\ref{fig:resids}.  The autocorrelation functions (ACFs) are shown for QI (red) and QIV (blue) and the cross correlation function between the two quadrants is shown in green.   
\label{fig:correlation}}
\end{figure}

To robustly estimate the characteristic width of the structure in the residuals 
on Figure~\ref{fig:resids}, we compute the autocorrelation functions of both quadrant's rotation curves.   These are shown
on Figure~\ref{fig:correlation} and labelled "acf".  To construct the autocorrelations, we first re-sample the
residuals to a constant step size of 5 pc in Galactic radius.
%measured along the locus of subcentral points ($\Delta s = $5 pc where $s = R_0 \ell$ with $\ell$ the longitude in radians).
All correlation functions are computed using the routine 
{\it scipy-signal-correlate} from the python numerical package {\it numpy},
which handles edge effects in a conservative way. But given that our
data covers a range of only $\sim$5 kpc, edge effects still lead to increasing
uncertainty for lags of more than about $\pm$2 kpc.  
We have compared the scipy correlation program with other algorithms, and
they give similar results for lags less than half the total interval 
%of $s$ which is 3 to 9.5 kpc (20\arcdeg$<l<64\arcdeg$).
of $R$ used for the correlation, which is 3.5 to 7.5 kpc. 

The half-width to half maximum of the autocorrelation functions for the
first and fourth quadrants are 290 and 210 pc, respectively.  Their half
widths to first zero crossing are  690 pc (QI) and 730 pc (QIV).
This shows that the residuals have generally quite broad structure, in 
spite of a few sharp variations that can be seen in Figure~\ref{fig:resids}.  That we observe a consistent scale size for velocity fluctuations in the rotation curve  that is much greater than the resolution of the data suggests that we are probing a characteristic size scale for dynamical effects.  This size scale is of the same  order as that often suggested for the energy injection scale of turbulence in the Milky Way, $\sim 150$ pc \citep[e.g.][]{chepurnov10}, and similar to the scale of the largest dynamical objects in the Galaxy, specifically potential energy injection sites of clustered supernovae \citep{norman96}.

 In addition to auto-correlation analysis, we can more robustly estimate the degree of similarity between the rotation residuals of the first and fourth quadrants with a cross-correlation analysis. The cross-correlation function between the residuals two quadrants shows a distinct peak near zero lag on Figure~\ref{fig:correlation} ("ccf").  
The peak has normalised correlation of 0.27 at $+75$ pc shift
(positive $\Delta r_G$ meaning that the QIV residuals are shifted to higher radius relative
to the QI residuals at the peak).   The non-zero normalised correlation coefficient confirms the visual impression that the two curves are very similar in their structure.  Furthermore, a radial shift between the two curves is what we expect if the large-scale structure is induced by spiral arms.  The magnitude of this shift, however, is surprisingly small.

The large sidelobes at $\pm$2 kpc
lag for  each of the cross-correlation function and  auto-correlation functions show that the residuals have structure 
that is roughly sinusoidal with wavelength about 2 kpc.   This corresponds to  the sawtooth pattern that is apparent by eye between $4.2 < R < 7$ kpc. It has long been assumed that these large departures from a smooth circular rotation are induced by the spiral pattern of the inner Galaxy \citep[e.g.][]{roberts69,burton71}.  Other grand design spirals like M81 the perturbations give sinusoidal departures from circular rotation \citep[e.g.][]{visser80} with amplitude (peak-to-peak) about 20 km s$^{-1}$.   
Unfortunately, interpretation of the structure in the residuals (Figure~\ref{fig:resids}) is complicated by the changing projection of the line of sight as we move along the locus of sub-central points and cross spiral arms at various angles. Furthermore, the velocity field predictions are different for linear and non-linear density wave theories \citep{yuan69,roberts69}.   The cross-correlation function and autocorrelation functions
presented in Figure~\ref{fig:correlation} will be useful constraints for hydrodynamical models of Galactic structure and dynamics \citep[e.g.][]{pettitt14}.

Underlying most of our analysis is the assumption that the measurement of the terminal velocity represents gas at the tangent point and probes the true rotation curve.   Of course, that premise is flawed slightly because of non-circular rotation due to streaming motions near spiral arms or the elliptical orbits around the central bar.  Given the observational indications that the Milky Way has a long bar, extending over a half-length of $R\sim 5$ kpc and an angle of $\sim 28 - 38$\arcdeg\ from the Sun-Galactic Center line \citep[e.g.][]{wegg15}, we expect non-axisymmetric effects in the inner Galaxy rotation curve.   
%The presence of a bar, with its elliptical orbits, affects the assumption of circular rotation, which underlies our tangent point analysis.   
Recently \citet{chemin15} explored this topic thoroughly using a hydrodynamical simulation of a barred Milky Way-like galaxy.  From the simulation they measured the velocity field and therefore both the true rotation curve and the rotation curve assuming the maximum velocity is found at the tangent point.  Their analysis found that for $4.8 \leq R\leq 8$ kpc\footnote{Galactocentric radii have been adjusted for our assumption of $R_0=8.5$ kpc.} the average tangent point derived curve is very close to the true Galactic rotation curve, however, in the range $3 \leq R \leq 4.8$ kpc, the tangent point analysis deviates by as much as $\sim 15$\% for bar angles of $20\arcdeg -30 \arcdeg$.  \citet{chemin15}  found for their fiducial bar of orientation $23\arcdeg$, QI tangent point velocities over-estimated the true azimuthal velocity and QIV tangent point values under-estimated the azimuthal velocities.  Similar results appear in the velocity field modelling by \citet{fux99} and seem to agree well with the residual rotation curves derived here. 
Because of the complexities in attributing measured terminal velocities to the tangent point the next steps for understanding the Milky Way rotation curve must be either comparison with simulations \citep[e.g.][]{englmaier99,fux99,chemin15} or surface density mass modelling \citep[e.g.][]{mcgaugh16}.

%----------------------------------------------------------
\section{Summary}
\label{sec:summary}
%----------------------------------------------------------
We have used the VLA Galactic Plane Survey of \HI\ and 21-cm continuum
emission \citep{stil06} to derive the terminal velocity curve of the first Galactic
quadrant with a sampling of 4\arcmin.  We find excellent consistency between the structure of the first quadrant \HI\ terminal velocity curve and the CO terminal velocity curve derived by \citet{clemens85}, but with a systematic offset in the \HI\ suggesting that the CO curve slightly over-estimates the terminal velocities.  Combining our first quadrant \HI\ data with the terminal velocity curve of
the fourth quadrant from the Southern Galactic Plane Survey \citep{mcgriff05} published
in our companion paper \citep{mcgriff07}, we have produced  the most uniform, high resolution,
terminal velocity curve for the Milky Way interior to the solar circle.  

As seen previously \citep{kerr62,blitz91} there is a clear asymmetry between the terminal velocity curves of the two quadrants.  We confirm that an outward motion of the LSR alone cannot explain the asymmetry \citep{blitz91}, highlighting the need to include a non-axisymmetric azimuthal velocity component for the inner Milky Way.  We find that the measured terminal velocities in two quadrants are almost identical over the galactic radius ranges, $3.8 < R < 4.2$ kpc and from $R>7$ kpc to the solar circle.  Over the intermediate radius range ($4 < R < 7$ kpc), the amplitudes of the two quadrants differ by $5$ to $10~{\rm km~s^{-1}}$, but they show remarkably similar structure.  

Using the rotation curves for first and fourth quadrants together we have performed a new joint fit to the Milky Way rotation curve over the range $3~{\rm kpc} < R< 8$ kpc.  This curve can be used for estimating kinematic distances.  This joint fit smooths over the streaming motions contained in the previous first quadrant fit by \citet{clemens85} and should be used in preference to that curve for inner Galaxy kinematic distances. We also compared our data with the flat and ``Universal'' spiral galaxy rotation curves from \citet{reid14}, finding both to be reasonable fits to the \HI\ rotation curve when calculated assuming $\Theta_0 = 240$ \kms\ and $R_0=8.34$ kpc, instead of the IAU recommended values for the Galactic constants. Finally, we compared the velocity residuals for the two quadrants after subtraction of the fitted rotation curve.  The residuals  show remarkably similar structure that is roughly sinusoidal with a wavelength of about 2 kpc.

With this high resolution, uniform terminal velocity curve we have overcome the remaining observational problems limiting our knowledge of the Milky Way terminal velocities.  There is no indication from our measurements that either higher resolution or higher sensitivity observations would change the \HI\ rotation curve derived here over the range $R=3 - 8 $ kpc.  Further progress on the Milky Way rotation curve will need to come from the compilation of these \HI\ values with other measurements such as on-going maser parallax \citep[e.g.][]{reid14} or upcoming stellar parallax from GAIA. Those measurements, together with our \HI\ terminal velocity curve can be used to improve the estimates of the surface density of the Milky Way disk and spiral structure.

%--------------------------------------------------------- 
\acknowledgements We are grateful to S.\ McGaugh whose encouragement
lead us to complete this long overdue companion paper and who provided us with an
electronic version of the Clemens data  and to M.\ Reid for comments on an earlier version of this manuscript.  N.\ M.\ M.-G.\ acknowledges the support of the Australian Research Council through Future Fellowship FT150100024.
%--------------------------------------------
% Bibliography
%--------------------------------------------
\bibliographystyle{apj}

%\bibliography{references} %~naomi/tex/references.bib, bibtex file 

\normalsize

%--------------------------------------------
% Tables
%--------------------------------------------
\clearpage

\begin{deluxetable}{lcc}
  \tabletypesize{\scriptsize} 
\tablecaption{Measured \HI\ terminal velocities, $v_{LSR}$, as a function of Galactic longitude, $l$, from the VGPS.  The third column gives the terminal velocities, $v_{mod}$, predicted by the linear rotation curve shown in Figure~\ref{fig:rot_fit}.  This table is published in its entirety in the electronic edition of the journal.
\label{tab:termvels}}
\tablewidth{0pt}
\tablehead{\colhead{$l$ (deg)} & \colhead{$v_{LSR}$ (\kms)} &  \colhead{$v_{mod}$ (\kms)}}
\startdata
18.412   &   134.17       &   137.98 \\ 
18.478   &   134.17       &   137.78 \\ 
18.543   &   133.86       &   137.58 \\ 
18.608   &   133.23       &   137.38 \\ 
18.672   &   132.73       &   137.19 \\ 
18.738   &   132.73       &   136.99 \\ 
18.802   &   131.91       &   136.80 \\ 
\enddata                   
\end{deluxetable}          
\end{document}